\documentclass{wow3}
\usepackage{amsmath}
\usepackage{amssymb}
\usepackage{graphicx}
\usepackage{mathrsfs}

\newcommand{\beqa}{\begin{equation}}
\newcommand{\eeqa}{\end{equation}}
\newcommand{\no}{\nonumber}
\newcommand{\be}{\begin{eqnarray}}
\newcommand{\en}{\end{eqnarray}}

\title[Growth of groups of wind generated waves]{Growth of groups of wind generated waves}
\author{F. Drullion$^{1}$ and S.G. Sajjadi$^{1,2}$}
\affiliation{$^{1}$\, Department of Mathematics, ERAU, Florida, USA.\\
$^{2}$\, Trinity College, University of Cambridge, UK.}

\begin{document}

\pagenumbering{arabic}
\setcounter{page}{128}

\maketitle

\begin{abstract}
In this paper we demonstrate numerical computations of turbulent wind blowing over group of waves that are growing in time. The numerical model adopted for the turbulence model is based on differential second-moment model that was adopted for growing idealized waves by Drullion \& Sajjadi (2014). The results obtained here demonstrate the formation of  cat's-eye which appear asymmetrically over the waves within a group.  
\end{abstract}

\vskip0.3in

\section{Introduction}
The question of growth and decay of wind generated waves in the ocean has been studied
extensively but the interactions and energy transfer between the water wave and the ambient
wind is still not fully understood. Most of the studies (experimental, numerical or analytical)
consider monochromatic or idealized waves profile. As it has been observed extensively that
waves actually travel in groups for which the profile changes as the wave is traveling. In this
study we are considering the influence of grouping on the growth of ocean waves induced by
wind. We are concentrating our efforts on the region around the height were the real part
of the complex wave speed is equal to the mean flow velocity. This region called ``critical
layer'' is at the center of Miles' (1957) theory and Lighthill's (1962) interpretation of growth waves.
In this region closed streamlines structures called ``cat's-eye'' are developed. The larger
these structures are, the more disturbance of the wind flow above the wave occurs. In some
previous work, e.g. Drullion \& Sajjadi (2014), a high-Reynolds number stress closure model over a 
moving idealized
wavy surface was used to show that their size and position are dependent on the wave age
and wave steepness, which is in accordance with direct numerical simulations of Sullivan 
{\em et al.}(2000). In this
study we use the same Reynolds stress model over different groups of waves to determine
the height of the critical layer and the overall shape and size.

\section{Growth of waves within a group}

In this paper we report computations of turbulent flow over two groups, in a frame of reference moving with the wave, namely \textit{group1} and \textit{group2}, for three different wave ages: $c_r/U_*=0.98$, $c_r/U_*=3.5$, and $c_r/U_*=7$. In the case of {\it group1} the group amplitude is kept fixed, i.e. non-growing waves. In contrast,  
we consider a case where the wave grows.

For \textit{group1} we adopt the following profile:
$$z=a[\cos(kx)+\cos(k_1 x)+\cos(k_2 x)]$$
where $a$ is the wave amplitude which we have chosen to be 0.0015 m, $k$ is the wave number, $k_1=1+\sqrt{2}ak$ and $k_2=1-\sqrt{2}ak$. In both wave groups the wavelength is $\lambda=0.0508$ m.

The profile for \textit{group2} is taken to be:
$$z=a(t)[\cos(kx)+0.3\cos(k_1 x)+0.3\cos(k_2 x)]$$
where am, the amplitude is set at 0.00165, $k_1$ and $k_2$ are the same as for \textit{group1}.

In our computations, the domain is taken to six wavelengths horizontally and two wavelengths vertically. The groups only extend over four wavelengths (from $x=0$ to $x=4\lambda$) and are surrounded by a flat surface. The latter ensures the periodicity in boundary conditions in the $x$-direction.
The condition $U=U_\lambda$ is imposed for the top boundary condition, and for the south boundary condition 
we prescribe the orbital velocity of the wave groups.

The orbital velocities for the \textit{group1} (for $0\leq x\leq 4\lambda$) is given by
\be 
u&=&-c_g ak[\cos(kx)+k_1 \cos(k_1x)+k_2 \cos(k_2 x)]-c_g\no\\
v&=&-c_g ak[\sin(kx)+k_1 \sin(k_1x)+k_2 \sin(k_2 x)]\no
\en
where $c_g$ is the group velocity. Similarly for the \textit{group2} the imposed orbital velocities is taken to be
\be
u&=&-c_g a(t)k[\cos(kx)+0.3 k_1 \cos(k_1x)+0.3 k_2 \cos(k_2 x)]-c_g\no\\
v&=&-c_g a(t)k[\sin(kx)+0.3 k_1 \sin(k_1x)+0.3 k_2 \sin(k_2 x)]\no
\en 
Note that, for the flat surfaces surrounding the group portion on the south boundary, namely when
$x<0$ and $x>4\lambda$, we impose the conditions
$u=-c_g$ and $v=0$.

For the growing wave groups, the computational mesh is regenerated every 50 time steps, where each time step consists of 500 iterations and is increased as the waves become steeper. The growth factor for each wave within the group is $e^{K c_i t}$, where $K$ can be taken to be $k$, $k_1$ or $k_2$ and $c_i=8c_ga/\lambda$.
\begin{figure}
  \centering
  \includegraphics[width=12cm]{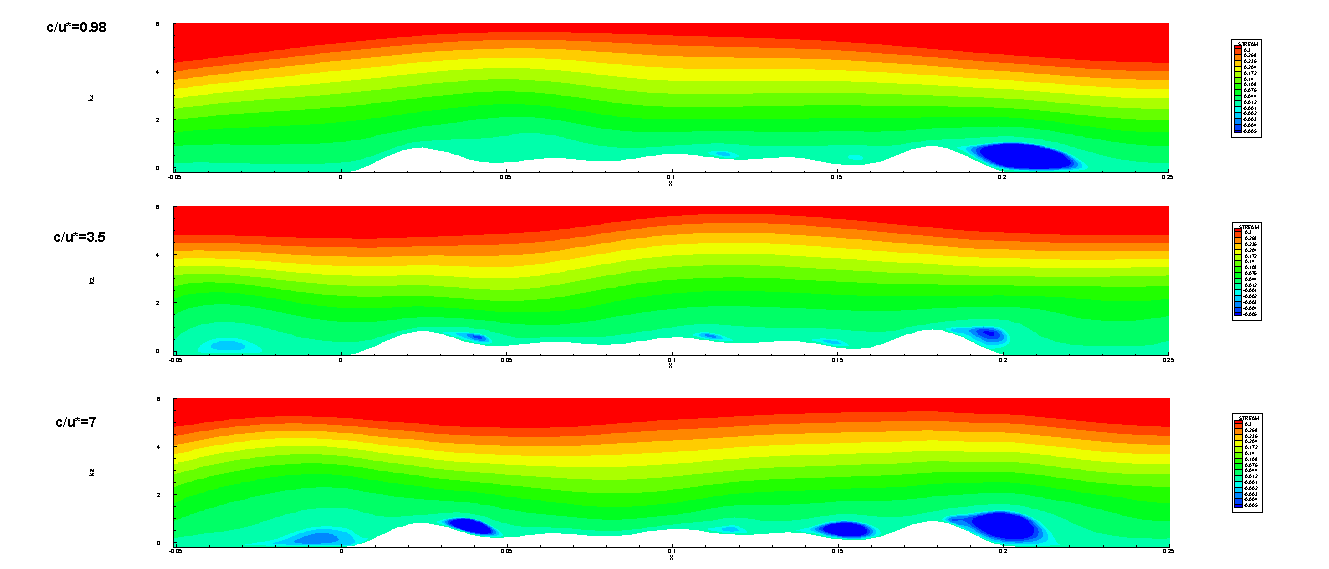}
  \caption{Stream function contour plots of stream function over {\it group1} wave for three values of the wave age $c_r/U_*=0.98$ (top); 3.7 (middle), and 7 (bottom).}
\label{fig:g1}
\end{figure}
\begin{figure}
  \centering
  \includegraphics[width=12cm]{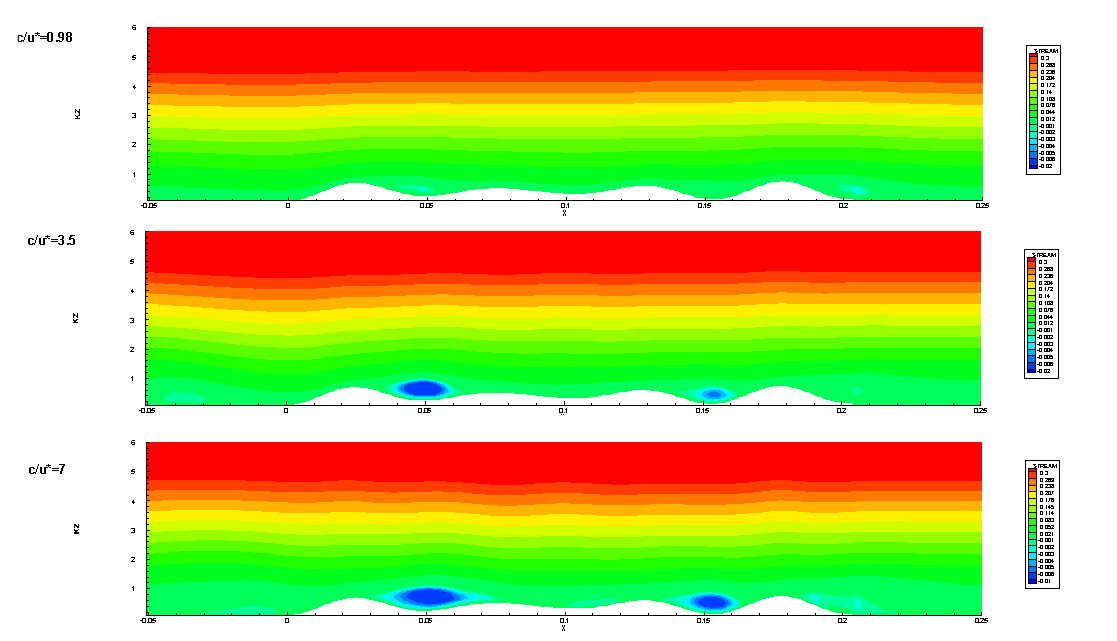}
  \caption{Stream function contour plots of stream function over {\it group2} wave for three values of the wave age $c_r/U_*=0.98$ (top); 3.7 (middle), and 7 (bottom). }
\label{fig:g2}
\end{figure}
\begin{figure}
  \centering
  \includegraphics[width=12cm]{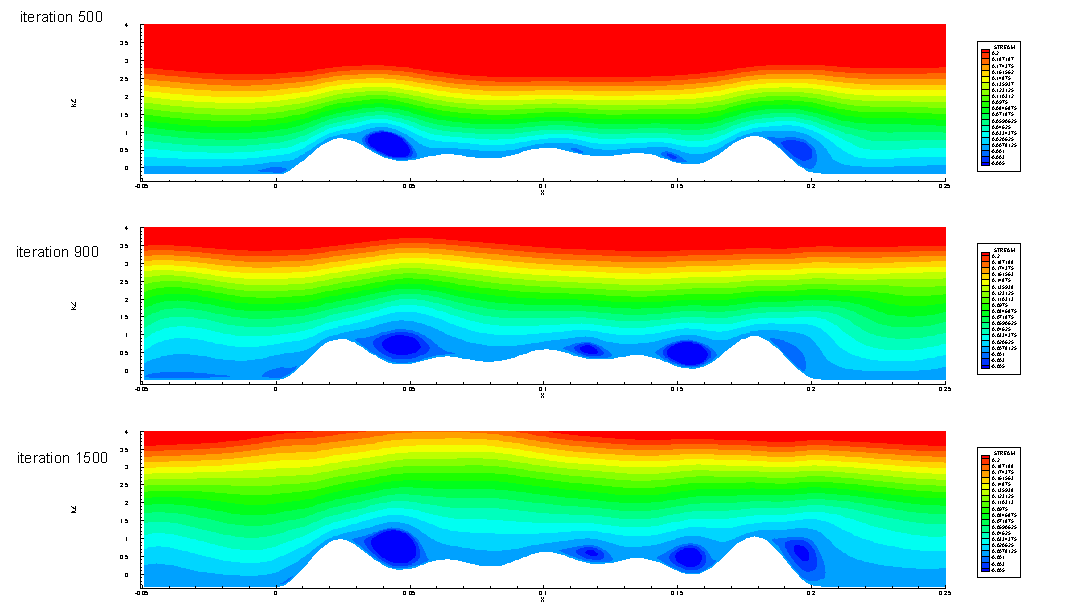}
  \caption{Stream function contour plots of stream function over growing {\it group1} wave for a single value of the wave age $c_r/U_*=5.75$. Sallower wave (top) and steepest wave (bottom).}
\label{fig:g1g}
\end{figure}

\section{Results and conclusions}

A computational turbulence model is used to compute the disturbed turbulent air flow over two different kind of wave group which we have referred to as {\it group1} and {\it group2}. 

Figure 1 shows the contour plot of the stream function for {\it group1} as a function for three wave ages $c_r/U_*= 0.98, 3.5$ and 7. As can be seen from this set of figures, at lowest value of $c_r/U_*$ a cat's-eye is formed downstream of the last wave in a group. As the wave age increases the same cat's eye become smaller but lifts up and moves toward the peak of the last wave, also formation of weaker cat's eye over other waves, within the group, become visible. At $c_r/U_*=7$ the cat's-eye become stronger and move further over the peak of the waves in the group. 

In figure 2 we perform the same computation for the {\it group2} for the same wave ages. In this case no cat's eye appear at $c_r/U_*=0.98$. However, at $c_r/U_*= 3.5$ two cat's eye is formed; a stronger one in the lee of the the first wave in the group, and a weaker one in the lee of the third wave in a group, but unlike {\it group1} no cat's-eye is formed in the lee of the last wave in the group. 

In figure 3 we show the the result of computations for a growing wave group {\it group1} for one fixed value of $c_r/U_*=5.75$. As can bee seen from these figures, as the wave steepens cat's-eye are formed in each lee of the wave in the group. As the wave grows so do the cat's-eyes, and similar to our other computations for monochromatic waves and bimodal Stokes waves (see also Sullivan {\em et al.} (2000)) the critical height rises further up from the surface of the waves. It is also evident that the flow become more asymmetrical 
which shows how the air flow over the downwind part of the group is lower
than over the upwind part. We remark that as the wave steepens the number of iteration must be increased in order to obtain a converged results. In what we have reported there is some doubt as the steepness is increased whether the solution for the steeper waves have completely converged. This is part of our on going investigation.    

We thus conclude that 
this asymmetry causes the critical layer height to be lower
over the downwind part. This is in line with the conclusion of our earlier paper, (Sajjadi, Hunt and Drullion (2016)) that the positive growth of the individual waves on the
upwind part of the wave group exceeds the negative growth on the downwind part. Hence, this leads to the
critical layer group effect producing a net horizontal force on the waves, in
addition to the sheltering effect.

\end{document}